\documentclass[a4paper,12pt]{article}
\usepackage{latexsym}
\usepackage{exscale}
\usepackage{graphicx}
\usepackage{amsmath}
\usepackage{amssymb}
\usepackage[germanpar]{anysize}
\marginsize{2.5cm}{2.5cm}{2.0cm}{2.0cm}
\begin{document}
%
%
\title{Numerically improved computational scheme for the optical
conductivity tensor in layered systems}
\author{A.~Vernes$^{\, a)}$, L.~Szunyogh$^{\, a,b)}$ and
P.~Weinberger$^{\, a)}$ \\
\ \\
$^{a)}$ Center for Computational Materials Science, \\
Technical University Vienna, \\
Gumpendorferstr. 1a, 1060 Vienna, Austria \\
$^{b)}$ Department of Theoretical Physics, \\
Budapest University of Technology and Economics \\
Budafoki \'{u}t 8, 1521 Budapest, Hungary}
\renewcommand{\today}{\textbf{submitted to J. Phys.:  Condensed Matter
(Sept. 19, 2000)}}
\maketitle
%
%
\begin{abstract}
The contour integration technique applied to calculate the optical
conductivity tensor at finite temperatures in the case of layered
systems within the framework of the spin--polarized relativistic
screened Korringa--Kohn--Rostoker band structure method is improved
from the computational point of view by applying the Gauss--Konrod
quadrature for the integrals along the different parts of the contour
and by designing a cumulative special points scheme for
two--dimensional Brillouin zone integrals corresponding to cubic
systems.
\end{abstract}
%
\section{Introduction}
%
\label{sect:intro}
Nowadays magneto-optical effects are widely used to probe the magnetic
properties of various systems \cite{BSE90,HTB97,SKB+98}. For a
theoretical description of these effects, one needs to calculate the
optical conductivity tensor in a parameter--free manner. Recently, two
of the authors have proposed a new, contour integration technique to
calculate the optical conductivity tensor for surface layered systems
\cite{SW99}. The theoretical framework of the present paper is based
on this technique. Therefore, in the following the basic concepts of
this method are only briefly reviewed.

The starting point of the contour integration technique is the
expression for the optical conductivity tensor
\begin{equation}
\Sigma_{\mu \nu} (\omega) = 
\dfrac{\sigma_{\mu \nu}(\omega)-\sigma_{\mu \nu}(0)}
{\hbar\omega + i\delta}
\label{eq:optcond}
\end{equation}
as given by Luttinger \cite{Lut67}, where $\omega$ denotes the photon
frequency and $\delta$ a finite life--time broadening,
respectively. The latter accounts for those scattering processes,
which are not incorporated in a standard band structure calculation,
but are present at finite temperatures.

\begin{figure}[htbp] \centering
\includegraphics[width=0.6\columnwidth,clip]{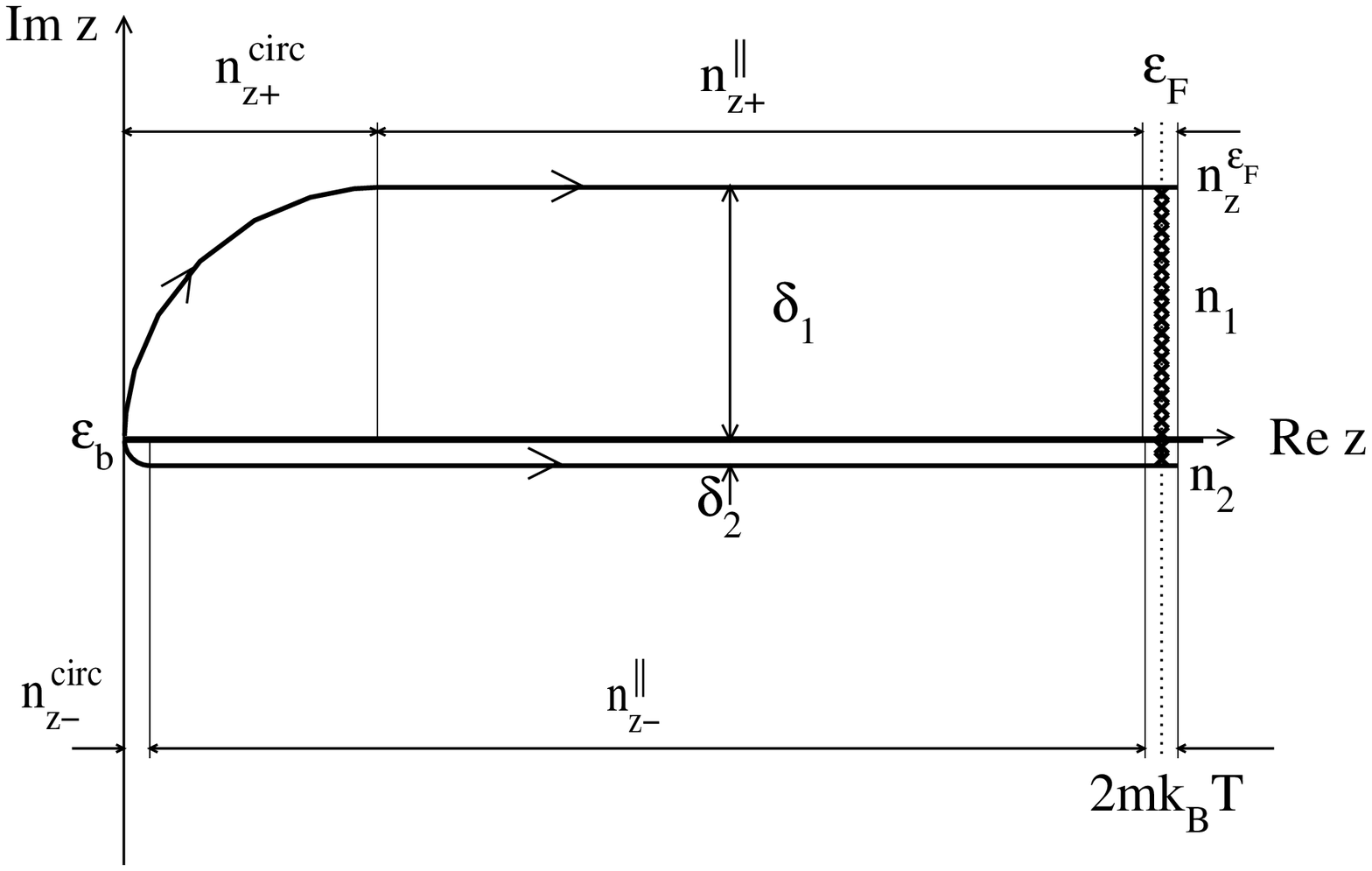}
\caption[Contour path and Matsubara poles.]
    {\label{fig:contour}
    Contour path and Matsubara poles used to calculate the zero
    wavenumber current--current correlation function. ($n_{z\pm}$
    denotes the complex energy points used on different parts of the contour. 
    $n_1$ and $n_2$ are Matsubara poles. $\epsilon_{\mathrm{b}}$ is the
    band bottom energy and $\epsilon_{\mathrm{F}}$ the Fermi level, 
    respectively.)
    }
 
\end{figure}

The temperature $T$ enters the expression for the zero wavenumber
current--current correlation function parametrically \cite{WC74}
\begin{equation}
\sigma_{\mu \nu} (\omega) = \dfrac{i\hbar}{V}\, \sum_{m,n}
\dfrac{f(\epsilon_n)-f(\epsilon_m)}
{\hbar\omega + i\delta + (\epsilon_n - \epsilon_m )} \,
J^{\mu}_{nm}J^{\nu}_{mn}
\label{eq:curcor}
\end{equation}
via the Fermi--Dirac distribution function $f(\epsilon)$, with
$\epsilon_m$ and $\epsilon_n$ being the eigenvalues of the
one--electron Hamiltonian corresponding to states labeled by $m$, $n$
and the $J^{\mu}_{nm}$ $(\mu = \mathrm{x,y})$ the current matrices.
In Ref.\ \cite{SW99} it was shown, that Eq.\ (\ref{eq:curcor}) can be
evaluated by means of a contour integration using complex energy
values $z$.  Within this technique, $\sigma_{\mu\nu}(\omega)$ is
decomposed into a contour path contribution
$\sigma^{(\mathrm{C})}_{\mu\nu}(\omega)$ and a contribution,
$\sigma^{(\mathrm{M})}_{\mu\nu}(\omega)$, arising from the Matsubara
poles $\sigma^{(\mathrm{M})}_{\mu\nu}(\omega)$, such that
\begin{equation}
\sigma_{\mu \nu} (\omega) = \sigma^{(\mathrm{C})}_{\mu\nu}(\omega) +
\sigma^{(\mathrm{M})}_{\mu\nu}(\omega) \, .
\label{eq:splitccf}
\end{equation}
As shown in Fig.\ \ref{fig:contour},
$\sigma^{(\mathrm{C})}_{\mu\nu}(\omega)$, in turn, consists of the
contributions from the contour in the upper and lower semi--plane as
given by Eqs.\ (24) and (25) in Ref.\ \cite{SW99}.  One contribution
to $\sigma^{(\mathrm{M})}_{\mu\nu}(\omega)$ comes from the $n_2$
Matsubara poles near and on both sides of the real axis, and an other
one from the $n_1$ poles situated exclusively in the upper
semi--plane, see Eq.\ (26) in Ref.\ \cite{SW99}. (Note that according
to Ref.\ \cite{SW99}, $n_1=N_1-N_2$ and $n_2=N_2$.)  Each of these
contributions (altogether four) is expressed in terms of
\begin{equation}
\tilde{\sigma}_{\mu \nu} (z_1,z_2) = - \dfrac{\hbar}{2\pi V} \,
\mathrm{Tr} \, \left[
J^{\mu} G(z_1) J^{\nu} G(z_2)
\right] \, ,
\label{eq:kubogreen}
\end{equation}
where $J^{\mu}$ and $G(z_i)$ denote current operators and resolvents,
respectively. Application of this contour integration technique to
compute $\Sigma_{\mu\nu}(\omega)$ for ordered or disordered (within
the framework of the single site Coherent Potential Approximation)
layered systems is therefore straightforward \cite{WLB+96}.
%
%
Originally, the quantities $\tilde{\sigma}_{\mu \nu} (z_1,z_2)$ were
introduced to facilitate the computation of the dc conductivity based
on the Kubo--Greenwood formalism \cite{Kub57,Gre58} in case of
substitutionally disordered bulk systems \cite{But85}.

In the present paper, the Green functions $G(z)$ entering Eq.\
(\ref{eq:kubogreen}) are calculated by means of the spin--polarized
relativistic screened Korringa--Kohn--Rostoker (SKKR) method for
layered systems \cite{SUWK94,SUW95,USW95} and the matrix elements
$J^{\mu}_{mn}$ ($\mu=\mathrm{x,y}$) using the relativistic current
operator \cite{WLB+96}. Because of the non--vanishing imaginary part
of the complex energies $z$ (see Fig.\ \ref{fig:contour}), also the
so--called irregular solutions of the Dirac equation have to be
considered \cite{SW99}.

To start a computation of the optical conductivity tensor
$\Sigma_{\mu\nu}(\omega)$ for a given frequency $\omega$ and
temperature $T$, the self--consistently determined potentials of the
investigated layered system must be known. This means that the bottom
energy $\epsilon_{\mathrm{b}}$ and the Fermi level
$\epsilon_{\mathrm{F}}$ are also known.  Thus $\min(\mathrm{Re}\, z)$
is fixed at $\epsilon_{\mathrm{b}}$ and $\max(\mathrm{Re}\, z)$ at
$\epsilon_{\mathrm{F}} + mk_{\mathrm{B}}T$ ($m\in\mathbb{N}$,
$k_{\mathrm{B}}$ the Boltzmann constant).  Due to the fast decay of
the Fermi--Dirac distribution function, at a given $T$ the computed
optical conductivity tensor depends only slightly on the used value of
$m$. The results given in the present paper were obtained with $m=8$.
As can be seen from Fig.\ \ref{fig:contour}, the value of
$\mathrm{Im}\, z$ is $\delta_1$ in the upper and $\delta_2$ in the
lower semi--plane along the contour part parallel to the real axis.
Taking, $\delta_j = 2 n_j \delta_T $ ($j$=1,2), with $\delta_T = \pi
k_{\mathrm{B}} T$, it is ensured that the paths parallel to the real
axis fall in--between two successive Matsubara poles \cite{SW99}.

In order to make use of the symmetry properties of the
$\tilde{\sigma}_{\mu \nu} (z_1,z_2)$ in Eq.\ (\ref{eq:kubogreen}), for
the life--time broadening a value of $ \delta = 2 \delta_2 $ has been
chosen.  An other advantage of this set--up is that for example at $T$
= 300 K and $\delta \approx$ 0.05 Ryd, one needs only $n_2 = 2$
Matsubara poles.

The computed value of the optical conductivity
$\Sigma_{\mu\nu}(\omega)$ depends strongly on the number of complex
energy points used on different parts of the contour in both
semi--planes: quarter circle, parallel to the real axis and in the
vicinity of the Fermi level.  Furthermore, the number of
$\Vec{k}$--points used to calculate the scattering path operator and
$\tilde{\sigma}_{\mu \nu} (z\pm\hbar\omega+i\delta,z)$ at a given
energy $z$, (see Eqs.\ (53) and (54) from Ref.\ \cite{WLB+96}), is of
crucial importance.

The aim of the present paper, is to describe efficient numerical
methods to control the accuracy of the contour and $\Vec{k}$--space
integrations without increasing the computational effort: in Section
\ref{sect:zint} Konrod's extension to the Gauss quadrature as a proper
numerical method is discussed; in Section \ref{sect:kint} a new,
cumulative special points method is presented to compute
two--dimensional $\Vec{k}$--space integrals with arbitrary high
precision.  Finally, the independence of $\Sigma_{\mu\nu}(\omega)$ on
the contour path in the upper semi--plane, i.e.\ on the $n_1$
Matsubara poles, is shown separately in Section \ref{sect:ndep}.
%
\section{Contour integration by means of the Konrod--Legendre rule}
%
\label{sect:zint}
The $n$--point Gauss--Legendre integration rule \cite{PFT+92} can be
used directly to compute $\sigma^{(\mathrm{C})}_{\mu\nu}(\omega)$ in
Eq.\ (\ref{eq:splitccf}) by transforming the nodes $x_k$ and their
weights $w_k$ ($k=1,\ldots,n$) according to the different contour
parts (Fig.\ \ref{fig:contour}). The nodes are roots of the Legendre
polynomials $P_n(x)$ corresponding to the weighting function $w(x)=1$,
$x\in (-1,1)$ and are usually computed using the Newton method
\cite{AS72}. For the weights $w_k$, one needs also to compute the
derivatives of the $P_n (x)$ with respect to the argument
$P_n^{\prime}(x_k)$.

One way to estimate the accuracy is to repeat the above procedure for
$n+1$ points and compare the results. This requires the evaluation of
the integrand for all the newly generated $n+1$ complex energy points,
because $P_{n+1}(x)$ has no common roots with $P_n(x)$ \cite{PFT+92},
which of course is computationally very inefficient.

In 1965 Konrod has proposed a method to overcome the above difficulty
by demonstrating that one can create a set of $2n+1$ nodes including
all the nodes of an $n$--point Gauss quadrature
\cite{Kon65}. Furthermore, he also showed that each of the additional
$n+1$ nodes falls in--between two nodes of the $n$--point Gauss
quadrature. Thus, once the integrand $f(x)$ is evaluated for each of
the $2n+1$ nodes, $\tilde{x}_k$, both the Gauss (denoted by
$\mathcal{G}_n$)
\begin{equation} 
\mathcal{G}_n f = \sum_{k=1}^{n} f(\tilde{x}_{2k}) w_{2k}
\label{eq:gauss}
\end{equation}
and the Konrod sum (denoted by $\mathcal{K}_{2n+1}$)
\begin{equation} 
\mathcal{K}_{2n+1} f = \sum_{k=1}^{2n+1} f(\tilde{x}_{k}) 
\tilde{w}_{k} \, ,
\label{eq:konrod}
\end{equation}
are available. Usually, the nodes $\tilde{x}_{k}$ and weights,
$\tilde{w}_{k}$ and $w_{2k}$, are obtained from the spectral
factorization of the associated Jacobi--Konrod (Jacobi--Gauss) matrix
\cite{CGG+00}. The Jacobi--Gauss matrix is formed easily knowing the
recursion coefficients of those monic, orthogonal polynomials (say
Legendre polynomials), which correspond to the weighting function (in
our case the identity) \cite{PFT+92}. But these coefficients fill only
partially the Jacobi--Konrod matrix. Hence, to form the Jacobi--Konrod
matrix is not a trivial task.

Laurie \cite{Lau97} was the first, who in 1997 developed an algorithm,
based on the mixed moments method in order to generate the
Jacobi--Konrod matrix for even $n$. Recently, Calvetti et
al. \cite{CGG+00} extended Laurie's ideas to odd $n$ using a
divide--and--conquer scheme.

We have implemented Laurie's algorithm \cite{Lau97} to calculate the
$2n+1$ recursion coefficients entering the Jacobi--Konrod matrix. The
initialization requires $3n/2$ ($n$ even) elements of the
Jacobi--Gauss matrix derived from a common set of monic, orthogonal
polynomials. These are obtained using the algorithm described in Ref.\
\cite{Gau94}.  The same Ref.\ \cite{Gau94} is then used to generate
the nodes and weights for the Konrod rule and the weights for the
Gauss rule with machine dependent accuracy, say $10^{-15}$.  In
contrast to the implementation of the Konrod rule described in Ref.\
\cite{Pie+83}, the present one works for arbitrary even $n$, i.e.\
$\mathcal{G}_n f$ can be compared with $\mathcal{K}_{2n+1} f$ not only
for some particular values of $n$.  The final result of the quadrature
is then that obtained by means of the Konrod rule.  Since
$\sigma^{(\mathrm{C})}_{\mu\nu}(\omega)$ is not a scalar quantity for
each pair of $(\mu,\nu)$, $\mathcal{G}_{n} \sigma_{\mu\nu}(\omega)
\equiv \sigma^{(n)}_{\mu\nu}(\omega)$ is compared separately with
$\mathcal{K}_{2n+1} \sigma_{\mu\nu}(\omega) \equiv
\sigma^{(2n+1)}_{\mu\nu}(\omega)$. The integrand along a particular
part of the contour, see Fig.\ \ref{fig:contour}, is said to be
converged if the maximum difference
\begin{equation}
\max \mid \sigma^{(2n+1)}_{\mu\nu}(\omega) -
\sigma^{(n)}_{\mu\nu}(\omega) \mid \; \leq \; \varepsilon_z
\quad (\mu,\nu=\mathrm{x,y}) \, ,
\label{eq:zconv}
\end{equation}
where $\varepsilon_z$ is an arbitrary small number, e.g.\ machine
accuracy ($10^{-15}$).

\begin{figure}[htbp] \centering
\begin{tabular}{cc}
\includegraphics[width=0.5\columnwidth,clip]{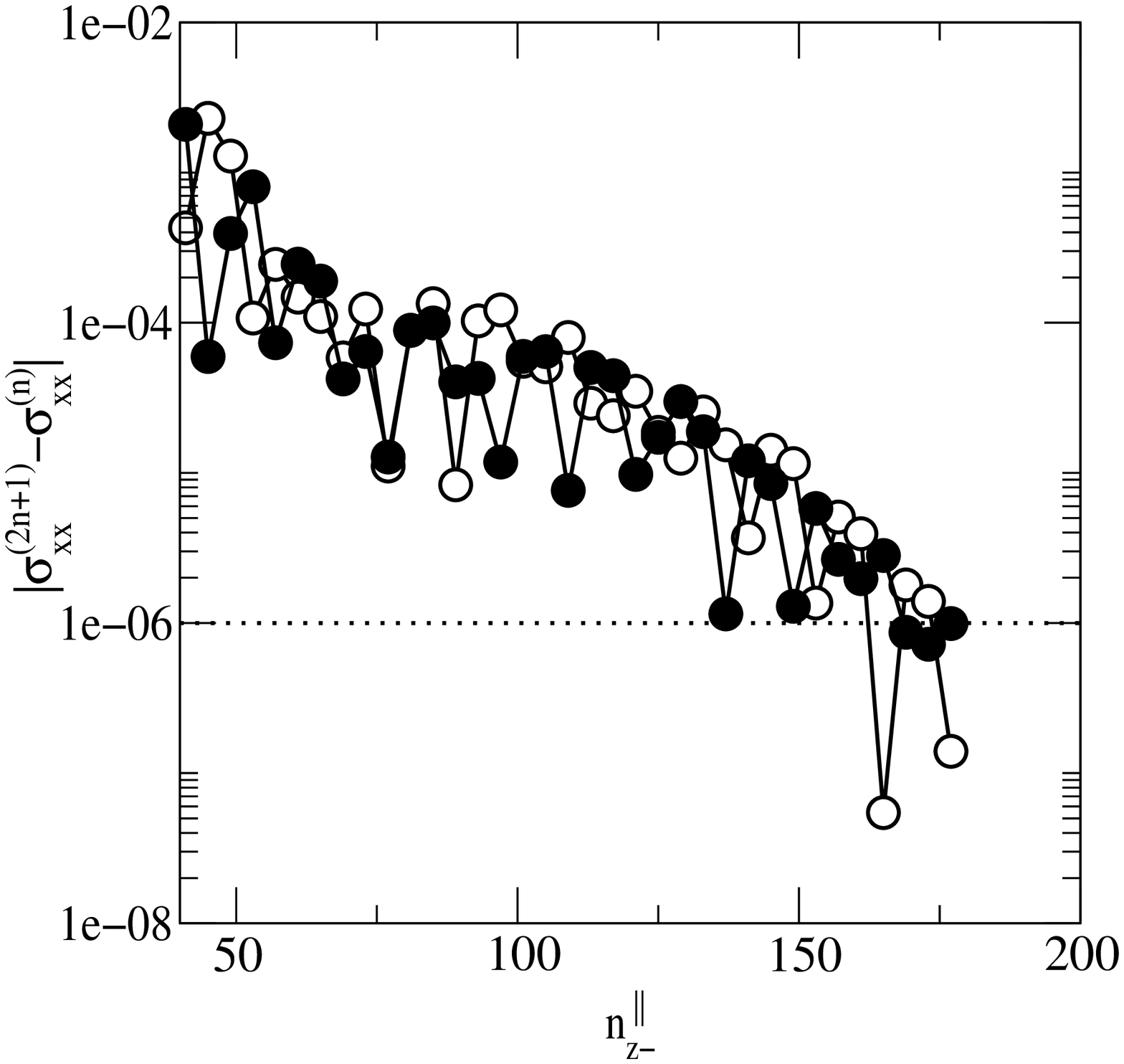} &
\includegraphics[width=0.5\columnwidth,clip]{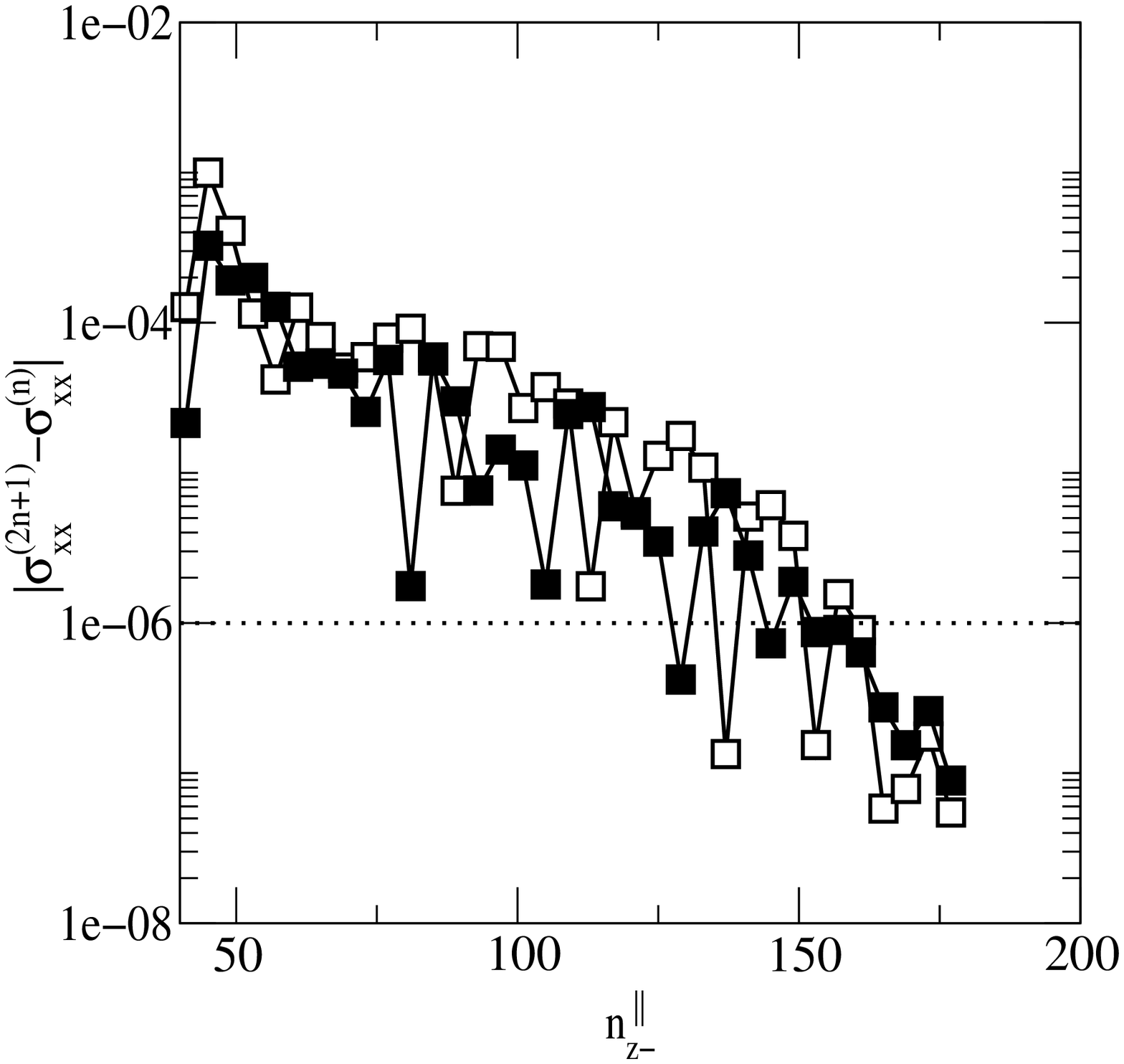} \\
\includegraphics[width=0.5\columnwidth,clip]{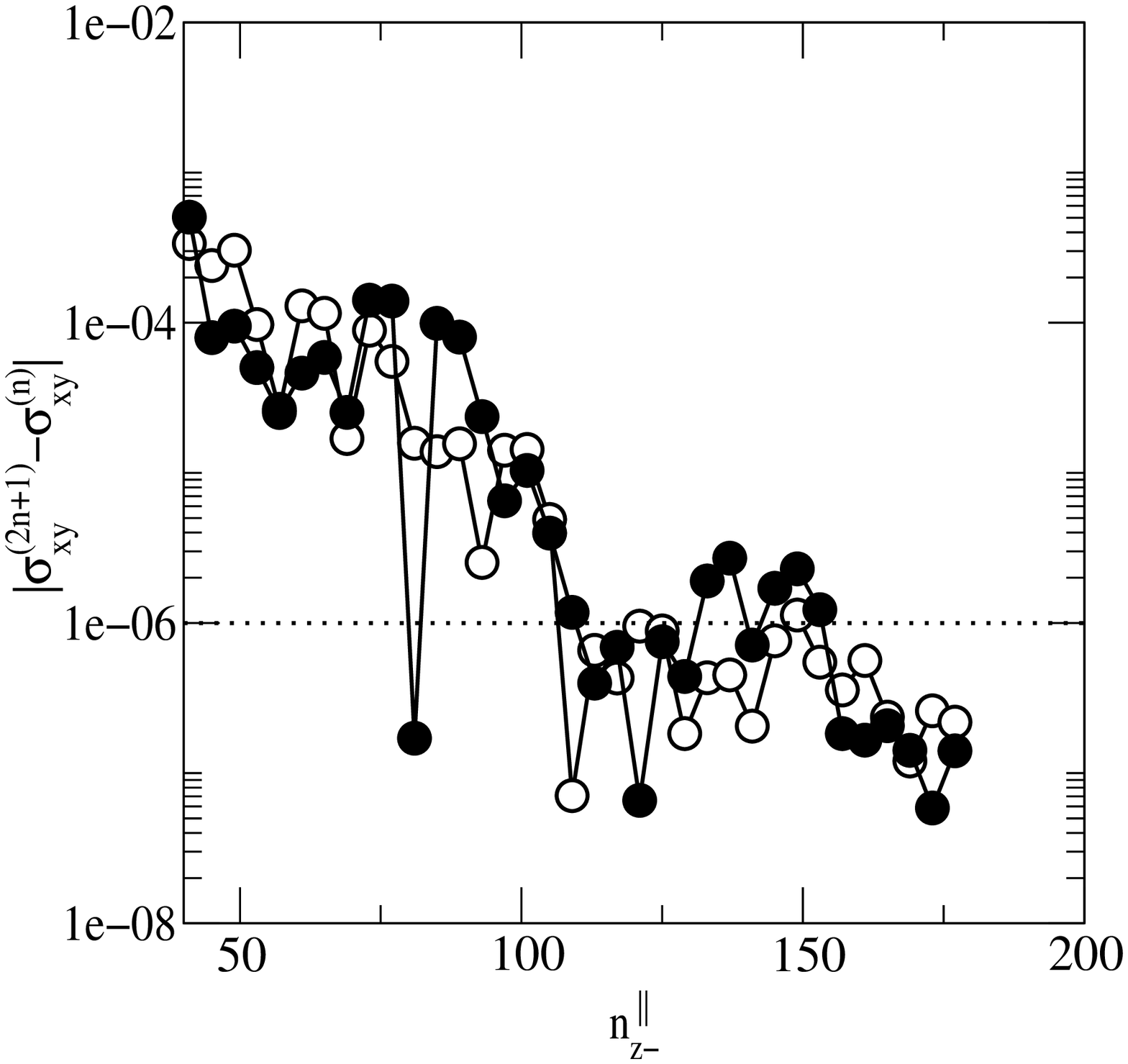} &
\includegraphics[width=0.5\columnwidth,clip]{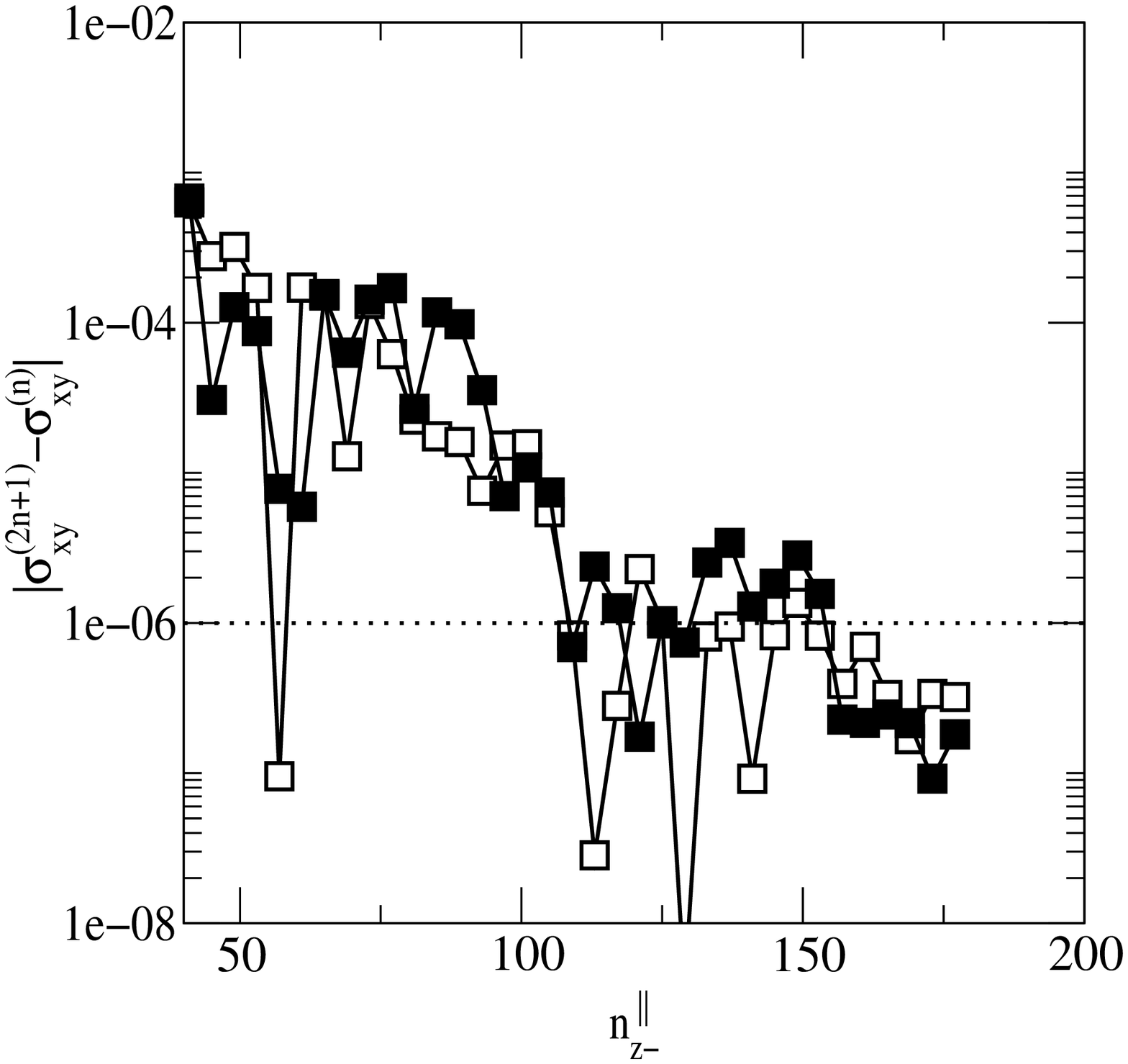}
\end{tabular}
\caption[Accuracy of the Gauss--Konrod rule in the lower semi--plane.]
    {\label{fig:zconv}
    Absolute difference between $\sigma_{\mu\nu}^{(2n+1)}(\omega)$
    and $\sigma_{\mu\nu}^{(n)}(\omega)$ (atomic units) versus the
    total number of complex energy points used parallel to the real axis in
    the lower semi--plane for $\hbar\omega$ = 0.05 Ryd and $T$ = 300 K. 
    The real part in the difference of the regular (left panels) and 
    irregular (right panels) contribution is given by open symbols 
    ($\circ$, $\Box$) and the imaginary part by corresponding filled
    symbols ($\bullet$, \footnotesize{$\blacksquare$}), \normalsize 
    respectively. 
    The dotted line marks a value of $10^{-6}$ in atomic units.)
    }
 
\end{figure}

In Fig.\ \ref{fig:zconv} $\mid \sigma^{(2n+1)}_{\mu\nu}(\omega) -
\sigma^{(n)}_{\mu\nu}(\omega) \mid$ is plotted for $\hbar\omega$ =
0.05 Ryd and $T$ = 300 K versus the number of complex energy points
$n_{z-}^{\parallel}$ used parallel to the real axis in the lower
semi--plane.
%
%
For all other parts of the contour, not shown here, the corresponding
quantities $\mid \sigma^{(2n+1)}_{\mu\nu}(\omega) -
\sigma^{(n)}_{\mu\nu}(\omega) \mid$, show an almost linear dependence
on $n$. The rather complicated shape of the data displayed in Fig.\
\ref{fig:zconv} has a physical reason: in the lower semi--plane, near
($\delta_2$ = 0.024 Ryd) and parallel to the real axis mostly the
joint density of states with all its singularities is mapped.

In the present paper, the layered system used for test calculations is
a mono--layer of Co on the top of fcc--Pt(100), i.e.\ below the Co
surface layer there are three Pt buffer layers followed by Pt bulk
\cite{PZU+99}. The band bottom energy ($\epsilon_{\mathrm{b}}$) and
the Fermi level ($\epsilon_{\mathrm{F}}$) of the substrate corresponds
to $-1$ and $-0.039$ Ryd, respectively. The optical conductivity
calculation was carried out using $n_1$ = 18 and $n_2$ = 2 Matsubara
poles. The number of $\Vec{k}$--points within the surface Brillouin
zone was 16 (further discussions on this point are made in Section
\ref{sect:kint}).  Analyzing the values obtained, it can be concluded
that for the case chosen, the following set of numerical parameters
\begin{displaymath}
\begin{array}[c]{rrrl}
n_{z+}^{\mathrm{circ}} = 4 &
n_{z+}^{\parallel} =\;\; 8 & 
n_{z+}^{\varepsilon_{\rm F}} = 6 &
\mbox{(upper semi--plane)} \\
n_{z-}^{\mathrm{circ}} = 4 &
n_{z-}^{\parallel} =  88   & 
n_{z-}^{\varepsilon_{\rm F}} = 8 &
\mbox{(lower semi--plane)} \\
\end{array}
\end{displaymath}
yields a maximum differences $\mid \sigma^{(2n+1)}_{\mu\nu}(\omega) -
\sigma^{(n)}_{\mu\nu}(\omega) \mid \; \leq 10^{-6}$ a.u. for each part
of the contour (dotted line in Fig.\ \ref{fig:zconv}). Notice that
$\varepsilon_z = 10^{-6}$ a.u.  is hundred times smaller than the
smallest contour part contribution to
$\sigma^{(\mathrm{C})}_{\mu\nu}(\omega)$.
%
\section{Cumulative special points method for two--dimensional
lattices}
%
\label{sect:kint}
In the present paper the special points method (SPM) has been
used. The reason for this is twofold. It can be shown that the SPM in
fact is a Gauss quadrature \cite{HaW92}. Therefore, its application
for a computation of $\Sigma_{\mu\nu}(\omega)$ guarantees that all
integrations involved are performed by means of the same quadrature
method. Furthermore, as demonstrated below, it is possible to use the
SPM cumulatively, which in turn facilitates to monitor the accuracy of
the $\Vec{k}$--space integration.

The integral $\mathcal{S}_{\mathrm{n}_{\Vec{k}}}$ of a function
$f(\Vec{k})$ over and normalized to the surface Brillouin zone (SBZ)
is approximated in the SPM similar to Eq.\ (\ref{eq:gauss}) by
\begin{equation} 
\mathcal{S}_{\mathrm{n}_{\Vec{k}}} f = 
\sum_{j=1}^{\mathrm{n}_{\Vec{k}}} f(\Vec{k}_j) w_j \, ,
\label{eq:spm}
\end{equation}
where $\mathrm{n}_{\Vec{k}}$ denotes the number of special points in
the irreducible part of the surface Brillouin zone (ISBZ), and the
weights $w_j$ have to fulfill the requirement:
\begin{equation} 
\sum_{j=1}^{\mathrm{n}_{\Vec{k}}} w_j = 1 \, .
\label{eq:sumw}
\end{equation}
These special points $\Vec{k}_j$ are defined by the following
condition
\begin{equation} 
\sum_{j=1}^{\mathrm{n}_{\Vec{k}}} w_j A_m(\Vec{k}_j) = 0 \, ,
\label{eq:spw}
\end{equation}
namely, in terms of a homogeneous system of linear equations in
symmetrized plane waves $A_m(\Vec{k})$ \cite{MF90,Lin78,MP76}, which
form a set of real, orthogonal, translationally and (point--symmetry
group) rotationally invariant functions \cite{CC73}.

Although there are several methods known in the literature to solve
Eq.\ (\ref{eq:spw}) for three-- \cite{MP76,MS92} and two--dimensional
\cite{Cun74} Brillouin zones, in the following we adopt the scheme
proposed by Hama and Watanabe \cite{HaW92}. They have shown that the
set of $\Vec{k}$--points
\begin{equation} 
\Vec{k}_j = 
\sum_{\alpha=1}^{2} k_{j_{\alpha}} \Vec{b}_{\alpha} \, ,
\label{eq:kmesh}
\end{equation}
with
\begin{equation} 
k_{j_{\alpha}} = \dfrac{j_{\alpha} - 1}{n_{\alpha}} 
+ \mathsf{a}_{\alpha} - \dfrac{1}{2}\, , \quad 
j_{\alpha}=1,\ldots,n_{\alpha}; \,\, \alpha=1,2 \,\, ,
\label{eq:kjalpha}
\end{equation}
are solutions of Eq.\ (\ref{eq:spw}), i.e.\ special points, which
minimize the remainder in Eq.\ (\ref{eq:spm}). Hence the special
points form an uniform, periodic mesh with respect to the edges
$\Vec{b}_{\alpha}$ ($\alpha=1,2$) of the reciprocal unit cell
\cite{MS92}, but they are not uniquely defined because of the
arbitrariness \cite{diffa} of the parameter $\mathsf{a}_{\alpha}$
($\alpha=1,2$) in Eq.\ (\ref{eq:kjalpha}).

The extension of the SPM proposed in the present paper exploits the
arbitrariness of $\mathsf{a}_{\alpha}$ and is based on the observation
that successively denser $\Vec{k}$--meshes, including all the
$\Vec{k}$--points of the previous meshes, can be created, if the
parameter $\mathsf{a}_{\alpha}$ in Eq.\ (\ref{eq:kjalpha}) does not
depend on $n_{\alpha}$. Consider a two times denser mesh
\begin{equation} 
k_{i_{\alpha}} = \dfrac{i_{\alpha} - 1}{2n_{\alpha}} 
+ \mathsf{a}_{\alpha} - \dfrac{1}{2}\, , \quad 
i_{\alpha}=1,\ldots,2n_{\alpha}; \,\, \alpha=1,2
\label{eq:kialpha}
\end{equation}
than that in Eq.\ (\ref{eq:kjalpha}). This new mesh includes \cite{mp}
all the (former) $\Vec{k}_j$--points and has additional points
in--between, because
\begin{equation} 
\left\{
\begin{aligned}
k_{i_{\alpha}} = & k_{j_{\alpha}}, \qquad\qquad \quad \mbox{if}\,\,
i_{\alpha}=2j_{\alpha}-1 \\ 
k_{i_{\alpha}} = & k_{j_{\alpha}} + \dfrac{\Delta k_{j_{\alpha}}}{2},  
\quad \,\, \mbox{if}\,\, i_{\alpha}=2j_{\alpha}
\end{aligned}
\right.
\label{eq:ki2kj}
\end{equation}
for $j_{\alpha}=1,\ldots,n_{\alpha}$ and
\begin{displaymath}
\dfrac{\Delta k_{j_{\alpha}}}{2} = \Delta k_{i_{\alpha}} =
k_{i_{\alpha}+1} - k_{i_{\alpha}} \, .
\end{displaymath}
It should be noted that the validity of the above statements does not
depend on the dimensionality of the Brillouin zone.

In our, cumulative SPM, we use origin centered $\Vec{k}$--meshes,
i.e.\ Eq.\ (\ref{eq:kjalpha}) for $\mathsf{a}_{\alpha}=1/2$ and the
same number of divisions $n_{\alpha}=n$ in each direction
$\alpha$. Since our interest in evaluating $\Sigma_{\mu\nu}(\omega)$
is mainly restricted to cubic layered systems, in Table\
\ref{tab:ockmesh} all details regarding the origin centered
$\Vec{k}$--meshes for primitive rectangular, square and hexagonal
lattices are listed.
%
 
\begin{table}[htbp] \centering
\begin{tabular}{|l|l|l|l|}\hline
 & & & \\
lattice & \multicolumn{1}{|c|}{$\Vec{k}_j\in\Omega_{\mathrm{ISBZ}}$, if} & 
\multicolumn{1}{|c|}{$w_j\,n_{\Vec{k}}(n)$} &
\multicolumn{1}{|c|}{$n_{\Vec{k}}(n)$} \\ 
 & & & \\ \hline
 & & & \\
primitive & $1\leq j_{\alpha}\leq \dfrac{n}{2} +1$ &
$4-2\left(\delta_{j_{1}1}+\delta_{j_{2}1}\right)+\delta_{j_{1}1}\delta_{j_{2}1}$
& $\left( n+1 \right)^{2}$ \\
rectangular & \multicolumn{1}{|r|}{($\alpha=1,2$)} & & \\ 
 & & & \\ \hline 
 & & & \\
square    & $1\leq j_{2}\leq j_{1} \leq \dfrac{n}{2} +1$ &
$4+4\left(1-\delta_{j_{1}j_{2}}\right)
\left(1-\delta_{j_{1}1}\right)\cdot$
& $\left( n+1 \right)^{2}$ \\
 & & $\left(1-\delta_{j_{2}1}\right)-3\delta_{j_{1}1}\delta_{j_{2}1}$ & \\ 
 & & & \\ \hline 
 & & & \\
hexagonal & $1\leq j_{1}\leq j_{2} \leq \dfrac{n}{2} +1$ &
$6-\left(2\delta_{j_{1}1}+3\right)\delta_{j_{2}1}$
& $1-9[m]\left( [m] +1 \right) $ \\
& $j_{1}+2j_{2} \leq n+3 $ & & $+3\left( \dfrac{n}{2}+1 \right)\left(
\dfrac{n}{2}+2[m]\right)$ \\ 
 & & & \\ \hline 
\end{tabular}
\caption[Origin centered 2D $\protect\Vec{k}$--meshes.]
    {\label{tab:ockmesh}
    Origin centered two--dimensional $\protect\Vec{k}$--meshes for
    even $n$ in the case of a primitive rectangular, square and 
    hexagonal lattice. $n_{\Vec{k}}(n)$ is the number of 
    points in the SBZ obtained by dividing the edges of the
    two--dimensional reciprocal unit cell into $n$ equal parts. 
    ($[m]$ denotes the integer part of $(n+2)/6$ and $\delta_{ij}$ 
    is the Kronecker symbol.)
}
\end{table}
%
 
As long as the magnetization is perpendicular to the surface, the
irreducible part of the SBZ is identical to the paramagnetic one.
(This situation pertains to the present paper.) The construction of
the paramagnetic ISBZ follows closely the one, introduced years ago by
Cunningham \cite{Cun74}. However, in the case of the hexagonal
lattice, a two times bigger ISBZ was taken, obtained by rotating
clockwise by 60$^{\circ}$ his ISBZ \cite{Cun74} and subsequent
mirroring along the $k_{\mathrm{x}}$ axis. The so obtained ISBZ and
$\Vec{k}$--meshes are in accordance with those of Hama and Watanabe
\cite{HaW92} used for the three--dimensional hexagonal lattice.

The weights $w_j$ in Table\ \ref{tab:ockmesh} were deduced using the
elements of the corresponding point--symmetry groups, i.e.\ ,
C$_{2\mathrm{v}}$ (primitive rectangular), C$_{4\mathrm{v}}$ (square),
and C$_{3\mathrm{v}}$ (hexagonal), respectively \cite{AH94}. They are
normalized to the total number of equivalent $\Vec{k}$--points in the
SBZ (last column of Table\ \ref{tab:ockmesh}) and fulfill the
condition in (\ref{eq:sumw}). It should be noted that all formulae in
Table\ \ref{tab:ockmesh} are valid only for even $n$.

When the cumulative SPM is used, 
\begin{equation}
\Delta n_{\Vec{k}}(n_i) = n_{\Vec{k}}(n_i) - n_{\Vec{k}}(n_{i-1})
\label{eq:dni}
\end{equation}
new $\Vec{k}$--points are added to a previous $\Vec{k}$--mesh ($i\geq
1$) and their contribution to the SBZ integral to be evaluated is
labelled by $\Delta\mathcal{S}_{n_i} f$.  If no previous mesh exists,
$n_{\Vec{k}}(n_{i-1})$ points are created according to Eqs.\
(\ref{eq:kmesh}) and (\ref{eq:kjalpha}) leading to the following
normalized sum
\begin{equation} 
\mathcal{S}_{n_{i-1}} f = \dfrac{1}{n_{\Vec{k}}(n_{i-1})} 
\sum_{j=1}^{n_{\Vec{k}}(n_{i-1})} f(\Vec{k}_j) \, .
\label{eq:bzspm}
\end{equation}
As a starting mesh $n_0=2m$ ($m$ arbitrary even number) can be
used. The subsequently created meshes then correspond to $n_i=2^i n_0$
($i=1,2,3,\ldots$) divisions along each vector $\Vec{b}_{\alpha}$.
Eq.\ (\ref{eq:bzspm}) with $\Delta n_{\Vec{k}}(n_i)$ newly created
$\Vec{k}$--points can also be used to compute $\Delta\mathcal{S}_{n_i}
f$.  Proceeding in this manner, one obtains a recursion relation of
type
\begin{equation}
\mathcal{S}_{n_i} f = \dfrac{1}{n_{\Vec{k}}(n_i)} 
\left[
\mathcal{S}_{n_{i-1}} f \cdot n_{\Vec{k}}(n_{i-1}) + 
\Delta\mathcal{S}_{n_i} f \cdot \Delta n_{\Vec{k}}(n_i) 
\right] \, .
\label{eq:cumspm}
\end{equation}
The $\Vec{k}$--points to be added to a previous mesh are selected in
terms of Eq.\ (\ref{eq:ki2kj}), by imposing that in Table\
\ref{tab:ockmesh} $j_1$ and $j_2$ cannot be simultaneously odd. It
should be noted that expressions to evaluate $\Delta n_{\Vec{k}}(n_i)$
directly can be also deduced from the $n_{\Vec{k}}(n)$ listed in
Table\ \ref{tab:ockmesh}. Eq.\ (\ref{eq:cumspm}) is then repeated
until the absolute difference between $\mathcal{S}_{n_i} f$ and
$\mathcal{S}_{n_{i-1}} f$ is smaller than a desired accuracy
$\varepsilon_{\Vec{k}}$ or an allowed maximum number of
$\Vec{k}$--points $n^{(\max)}_{\Vec{k}}$ is reached. In particular,
for $\Sigma_{\mu\nu}(\omega)$, this means that
\begin{equation}
\max \mid \sigma^{(n_i)}_{\mu\nu}(\omega) -
\sigma^{(n_{i-1})}_{\mu\nu}(\omega) \mid \; \leq \; \varepsilon_{\Vec{k}}
\quad (\mu,\nu=\mathrm{x,y}) \, .
\label{eq:kconv}
\end{equation}
is imposed for each complex energy on the contour and Matsubara
pole.

It should be noted that Eq.\ (\ref{eq:bzspm}) applies to the full SBZ,
whereas Eq.\ (\ref{eq:spm}) refers to an irreducible wedge of the SBZ
\cite{SUWK94,HW92}.

\begin{figure}[htbp] \centering
\begin{tabular}{cc}
\includegraphics[width=0.5\columnwidth,clip]{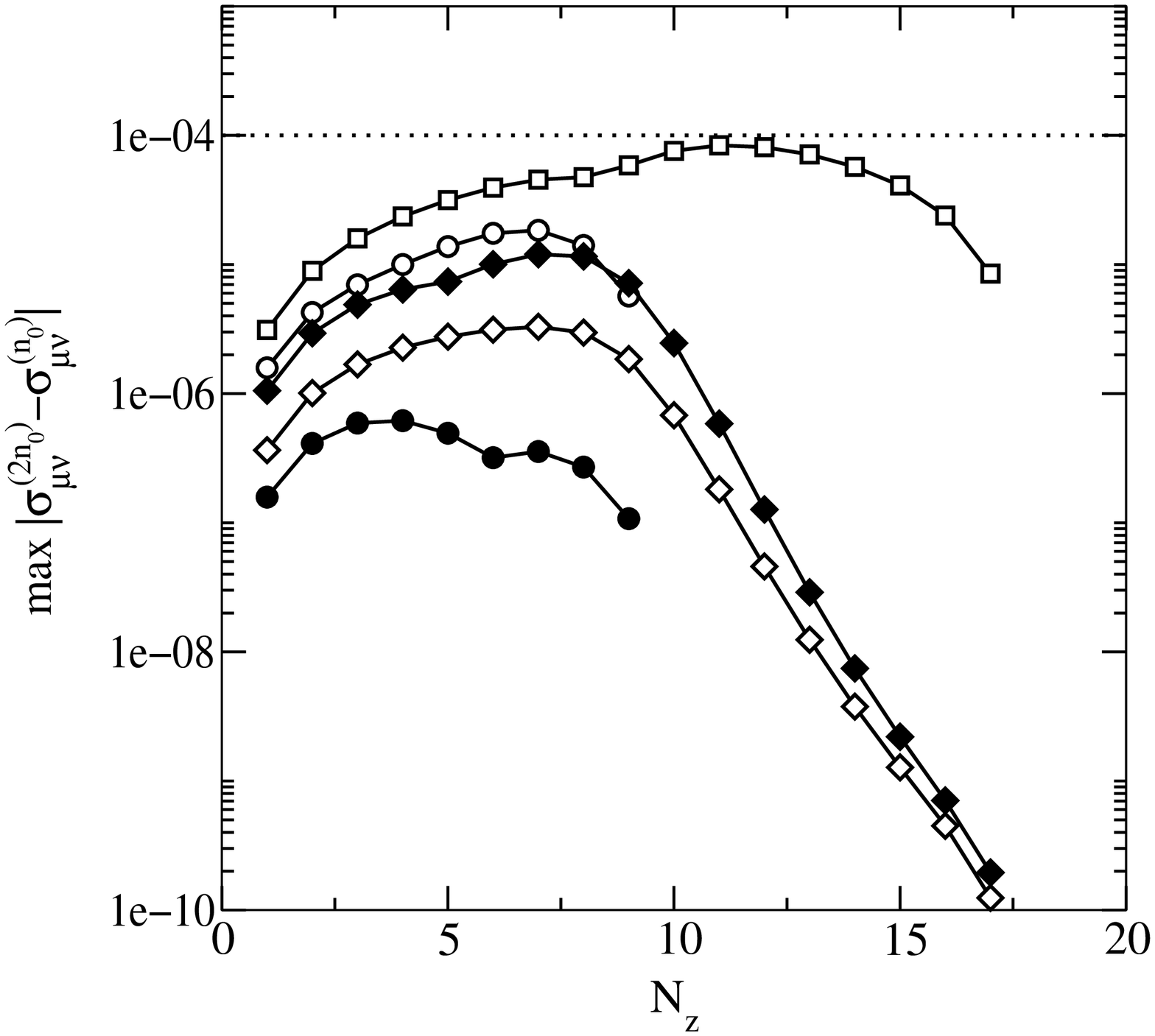} &
\includegraphics[width=0.5\columnwidth,clip]{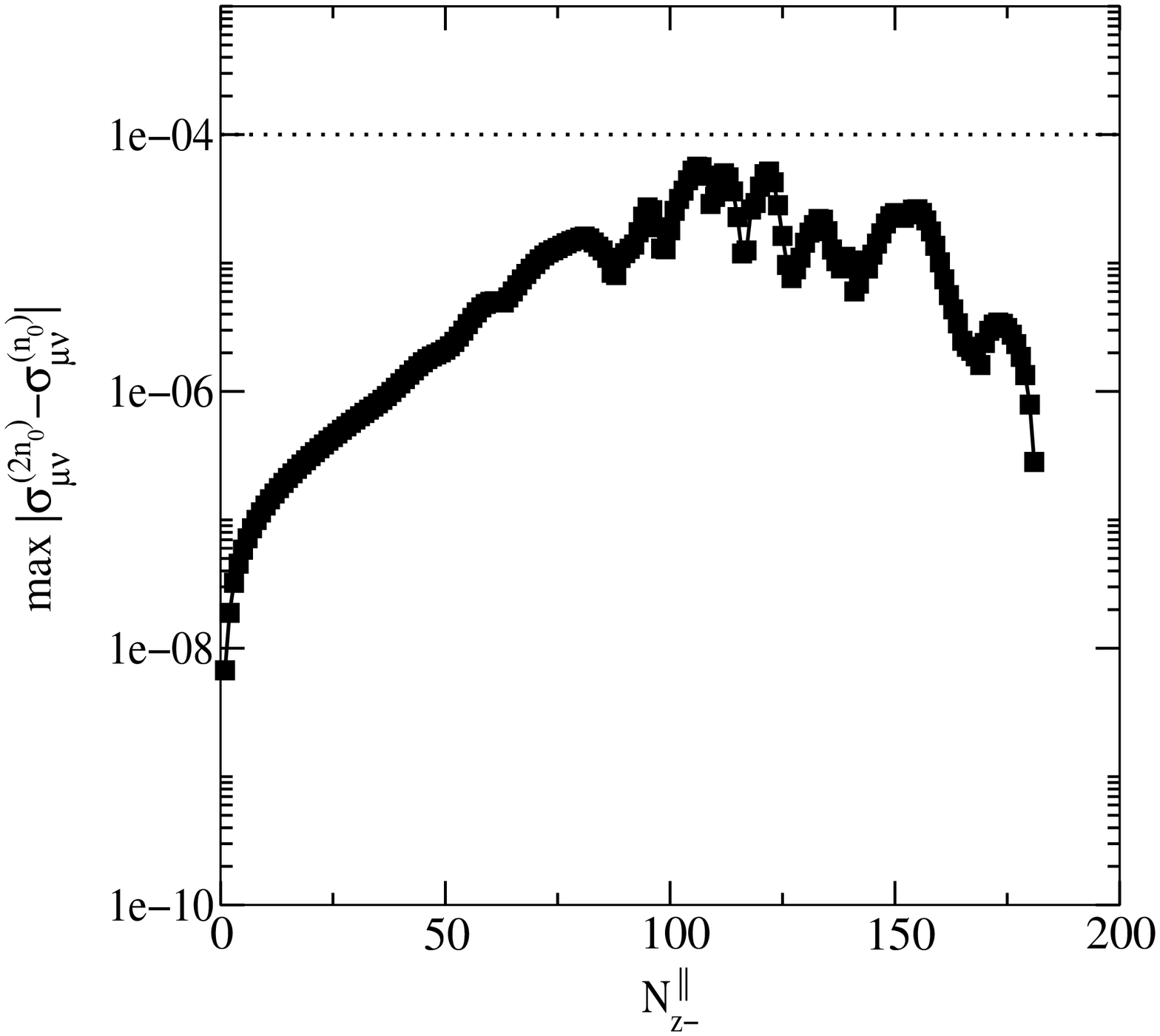} \\
\multicolumn{2}{c}{
\includegraphics[width=0.5\columnwidth,clip]{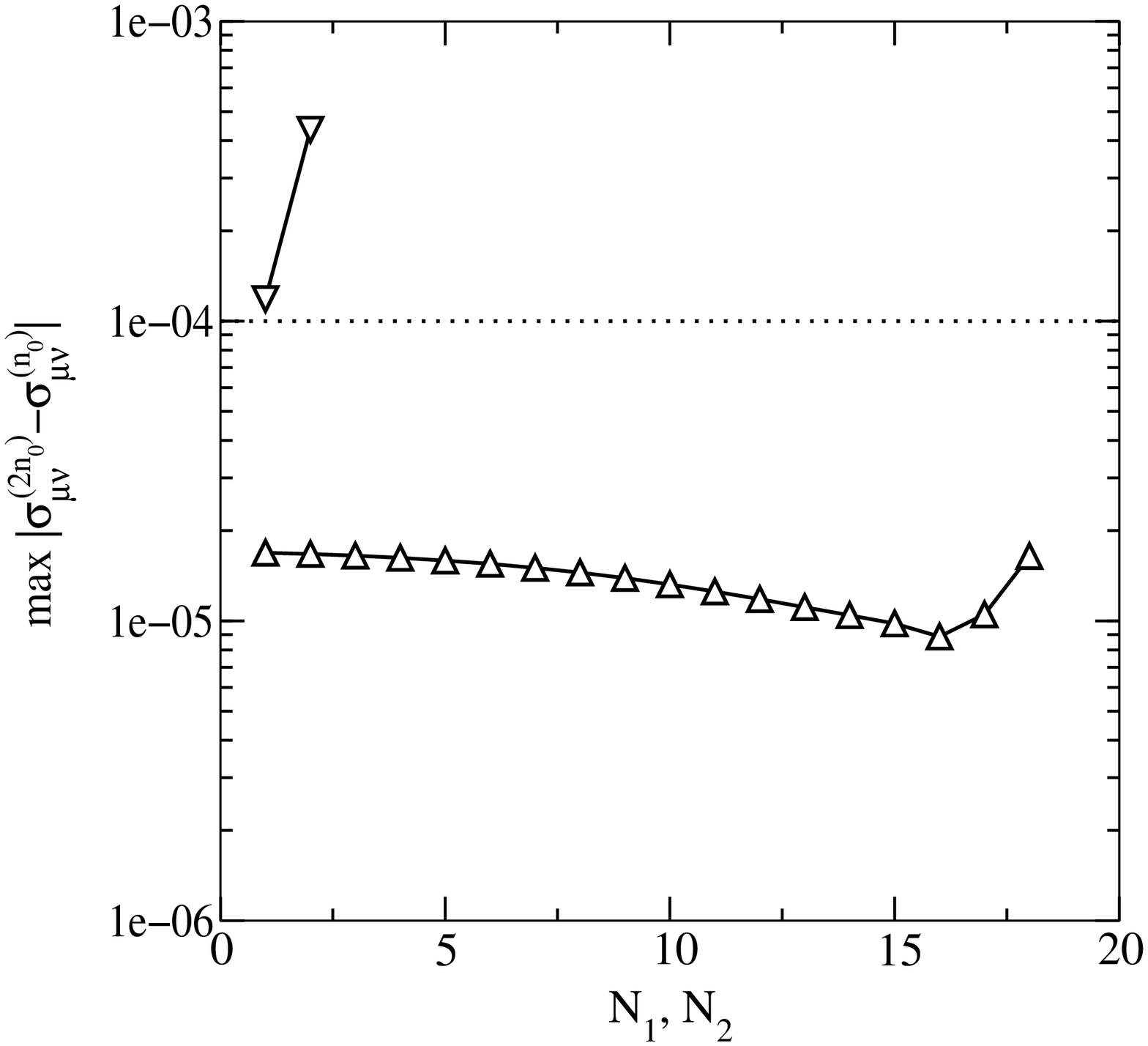} 
}
\end{tabular}
\caption[Accuracy of the cumulative special points method.]
    {\label{fig:kconv}
    Maximum absolute difference between $\sigma_{\mu\nu}^{(2n_0)}(\omega)$
    and $\sigma_{\mu\nu}^{(n_0)}(\omega)$ in atomic units for a 
    complex energy points $N_z$ on a part of the contour or for a Matsubara
    pole $N_1$, $N_2$ ($n_0=8$, $\hbar\omega$ = 0.05 Ryd and $T$ = 300 K). 
    Open (filled) symbols are used for the contour in the upper
    (lower) semi--plane. Circles refer to the
    quarter circles, squares to the parts parallel to the real axis
    and diamonds in the vicinity of the Fermi. $\vartriangle$ denotes the
    Matsubara poles far off from the real axis in the upper semi--plane
    and $\triangledown$ the others. The dotted line marks the value of
    $10^{-4}$ in atomic units.
    }
 
\end{figure}

An application of the cumulative SPM is shown in Fig.\
\ref{fig:kconv}. For these calculations ($\hbar\omega$ = 0.05 Ryd and
$T$=300 K), the same layered system is considered as in Section
\ref{sect:zint}. The results obtained with a starting $\Vec{k}$--mesh
consisting on 15 $\Vec{k}$--points in ISBZ ($n_{i-1}\equiv n_0=8$) is
taken as reference. In Fig.\ \ref{fig:kconv} these data are compared
with those obtained using 45 $\Vec{k}$--points in ISBZ
($n_i=2n_0=16$).  For the contour integrations an accuracy
$\varepsilon_z = 10^{-4}$ a.u. was achieved on each contour part (see
Fig.\ \ref{fig:contour} and Section \ref{sect:zint}).  This means that
even the minimum of $\tilde{\sigma}_{\mu\nu}(z_1,z_2)$ has a last
digit exactly computed.

As can be seen in Fig.\ \ref{fig:kconv}, a common precision of
$\varepsilon_{\Vec{k}}=10^{-4}$ a.u. can be achieved easily with a
one--step cumulative SPM for all parts of the contour and for the
Matsubara poles situated in the upper semi--plane far off from the
real axis. Obviously, for the Matsubara poles near the real axis more
$\Vec{k}$--points are needed in order to achieve the same accuracy
$\varepsilon_{\Vec{k}}$.
%
%
 
%
\section{Contour path independence}
\begin{figure}[htbp] \centering
\begin{tabular}{c}
\includegraphics[width=0.6\columnwidth,clip]{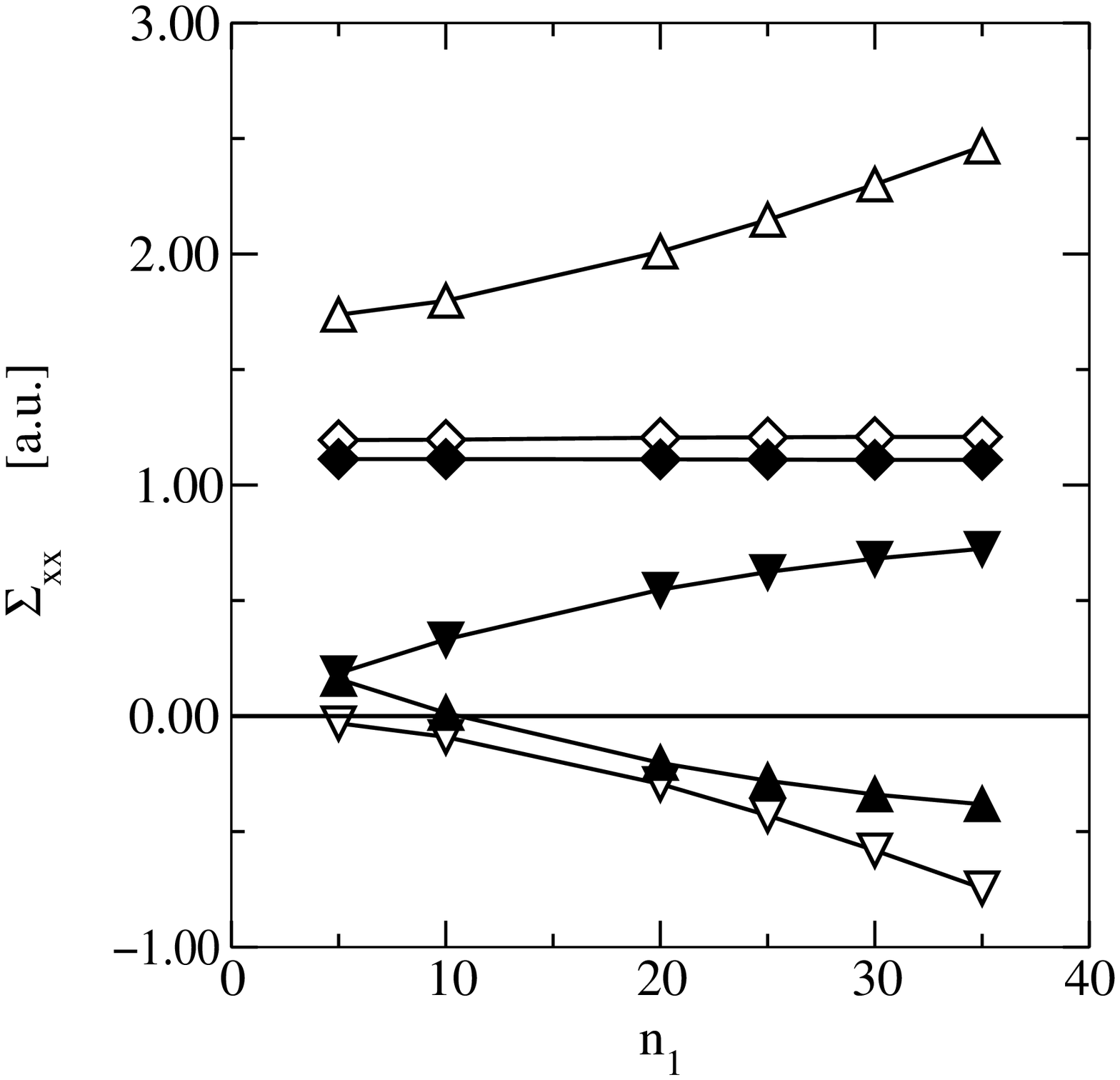} \\
\includegraphics[width=0.6\columnwidth,clip]{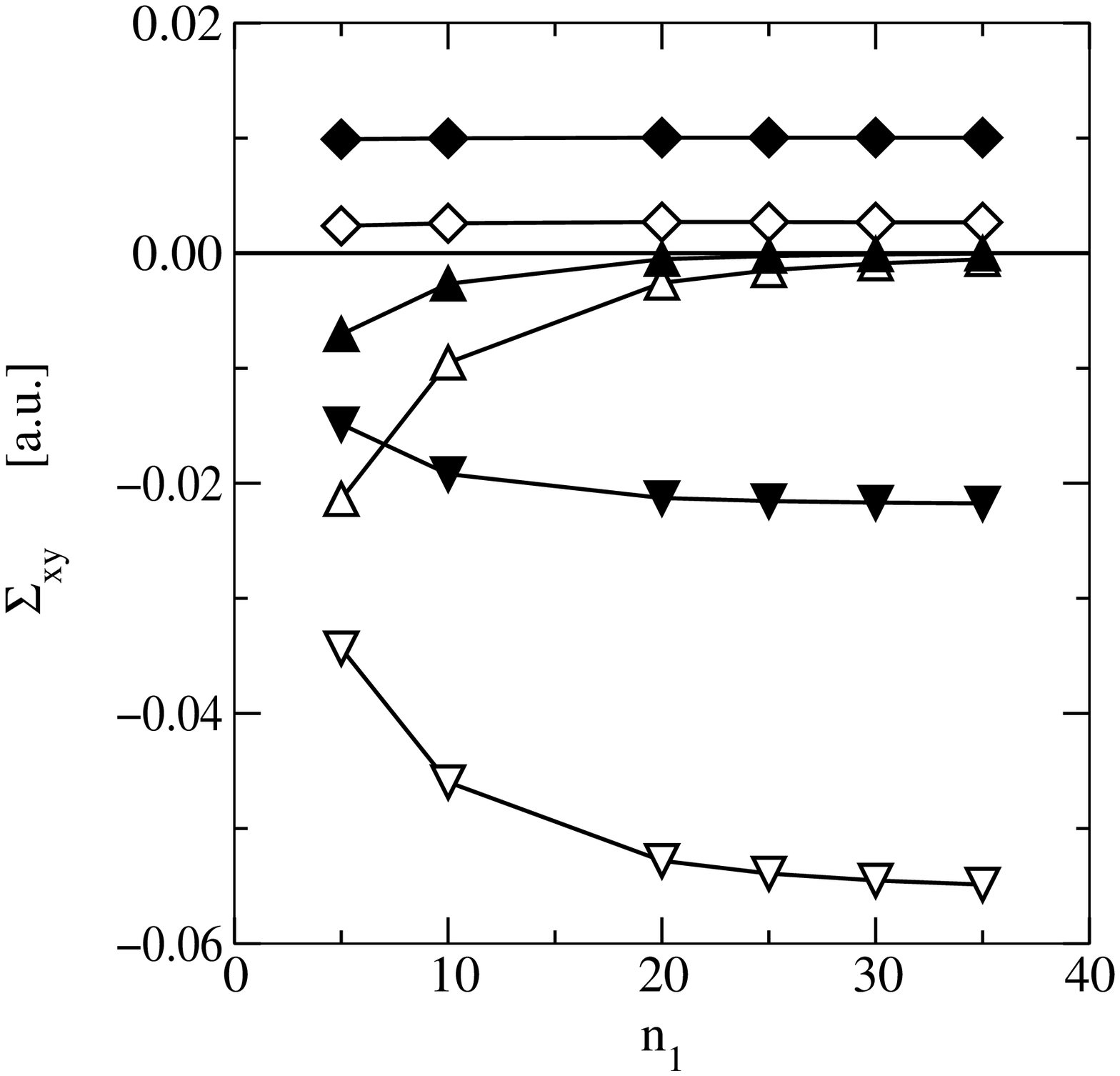} \\
\end{tabular}
\caption[Contour path independence in the upper semi--plane.]
    {\label{fig:ndep}
    Real (open symbols) and imaginary part (filled symbols) of
    $\Sigma_{\mu\nu}(\omega)$, respectively, in atomic units for 
    $\hbar\omega$ = 0.05 Ryd and $T$ = 300 K against the Matsubara 
    poles $n_1$ in the upper semi--plane. (Triangles up are used for the 
    contribution coming from the contour and triangles down for that
    coming from the Matsubara poles. Their sum is given by diamond. 
    The convergence criteria (\ref{eq:zconv}) and (\ref{eq:kconv}) for
    $\varepsilon_z = \varepsilon_{\Vec{k}}=10^{-4}$ 
    a.u. were satisfied for each $n_1$ value considered.)
    }
 
\end{figure}

%
\label{sect:ndep}
For the layered system described in Section \ref{sect:zint} in Fig.\
\ref{fig:ndep} the optical conductivity $\Sigma_{\mu\nu}(\omega)$
[a. u.] for $\hbar\omega$ = 0.05 Ryd and $T$ = 300 K is shown as
function of the Matsubara poles $n_1$ used in the upper semi--plane
far off from the real axis.  ($n_2=2$ Matsubara poles were used near
the real axis in both semi--planes.)  The convergence criteria
(\ref{eq:zconv}) and (\ref{eq:kconv}) were satisfied for
$\varepsilon_z=\varepsilon_{\Vec{k}}=10^{-4}$ [a.u.].

As can be seen from Fig.\ \ref{fig:ndep}, the contribution coming from
the contour $\Sigma^{(1)}_{\mu\nu}(\omega)$ and from the Matsubara
poles $\Sigma^{(3)}_{\mu\nu}(\omega)$, see also Eqs.\ (24) and (26) in
Ref.\ \cite{SW99}, in the upper semi--plane, respectively, depends
remarkably on $n_1$. However, their sum does not really depend on
$n_1$, i.e.\
\begin{displaymath}
\Sigma^{(1)}_{\mu\nu}(\omega) + \Sigma^{(3)}_{\mu\nu}(\omega) =
\mathrm{const.} \quad (\mu,\nu=\mathrm{x,y})
\end{displaymath}
with an accuracy of $n_1\varepsilon_{\Vec{k}}\approx 10^{-3}$ [a.u.].
Hence an evaluation of $\Sigma_{\mu\nu}(\omega)$ does not depend on
the form of the contour in the upper semi--plane.
%
\section{Summary}
%
The computational scheme for the optical conductivity tensor for
layered systems has been improved numerically. For the contour
integration, the Konrod--Legendre rule was applied, showing that any
desired accuracy $\varepsilon_z$ can be achieved (in comparison with
the Gauss--Legendre rule). In the case of the $\Vec{k}$--space
integration, a cumulative special points scheme was developed for
two--dimensional lattices. This method permits one to perform for each
complex energy $z$ the $\Vec{k}$--space integration iteratively,
evaluating the integrand only for those $\Vec{k}$--points added to a
previous mesh.  The thus controlled $z$-- and $\Vec{k}$--convergence,
provides independence from the form of the contour in the upper
semi--plane with a predictable accuracy.

It should be noted that the described numerical procedures can be used
also for other approaches or to calculate other physical
properties. For example, the cumulative special points method provides
an excellent tool to check the $\Vec{k}$--convergence of the band
energy part of the magnetic anisotropy energy. The numerical
efficiency in calculating the transport properties can be improved in
a similar way.
%
\section{Acknowledgements}
%
This work was supported by the Austrian Ministry of Science (Contract
No. 45.451/1-III/B/8a/99) and by the Research and Technological
Cooperation Project between Austria and Hungary (Contract
No. A-35/98).  One of the authors (L.S.) is also indebted to partial
support by the Hungarian National Science Foundation (Contract No.
OTKA T030240).
%
%

%
%
\newpage
%
 
%
%
 
%
%
 
%
%
 
%


\begin{thebibliography}{10}
\bibitem{diffa} 
$\Vec{k}$--meshes, for example, with $\mathsf{a}_{\alpha}=0,
1/n_{\alpha}$ ($\alpha=1,2,3$) are given in Ref.\
\cite{HWK90}. $\mathsf{a}_{\alpha}=1/2$ was applied to the
three--dimensional hexagonal lattice in Ref.\ \cite{HaW92}. The
frequently used, so--called Monkhorst--Pack meshes \cite{MP76} are
obtained considering $\mathsf{a}_{\alpha}=1/2n_{\alpha}$. Particular
two--dimensional Monkhorst--Pack meshes are given in Ref.\
\cite{Cun74}.
\bibitem{mp}
As a corollary, it can be shown that a Monkhorst--Pack mesh cannot
include all the $\Vec{k}$--points of an other, similar Monkhorst--Pack
mesh.
\end{thebibliography}
\end{document}